\documentclass[prl,aps,twocolumn,noshowpacs,a4paper]{revtex4}
\usepackage{bm}
\usepackage[dvips]{graphicx}
\usepackage{amssymb,amsfonts,amsmath,color}

\newcommand{\be}{\begin{equation}}
\newcommand{\ee}{\end{equation}}
\newcommand{\bea}{\begin{eqnarray}}
\newcommand{\eea}{\end{eqnarray}}
\newcommand{\bse}{\begin{subequations}}
\newcommand{\ese}{\end{subequations}}
\newcommand{\bc}{\begin{cases}}
\newcommand{\ec}{\end{cases}}

\newcommand{\bs}{\boldsymbol}

\newcommand{\mrm}{\mathrm}

\newcommand{\bpm}{\begin{pmatrix}}
\newcommand{\epm}{\end{pmatrix}}

\newcommand{\go}{\omega}
\newcommand{\gt}{\theta}

\newcommand{\Gc}{\Gamma}

\newcommand{\Gd}{\Delta}

\newcommand{\Gl}{\Lambda}

\newcommand{\diff}{\mrm{d}}

\newcommand{\lan}{\langle}
\newcommand{\ran}{\rangle}

\newcommand{\met}{\mrm{m}}

\renewcommand{\sec}{\mrm{s}}

\begin{document}

\title{Fluid dynamics and noise in bacterial cell-cell and cell-surface scattering}

\author{
Knut Drescher$^{1}$, 
J{\"o}rn Dunkel$^{1}$, 
Luis H. Cisneros$^{2}$, 
Sujoy Ganguly$^{1}$, 
Raymond E. Goldstein$^{1}$}
\affiliation{$^{1}$Department of Applied Mathematics and Theoretical Physics, University of Cambridge, Wilberforce Road, Cambridge CB3 0WA, UK}
\affiliation{$^{2}$Department of Physics, University of Arizona, 1118 E 4th St, Tucson, Arizona 85721, USA}


\begin{abstract}
\noindent
Bacterial processes ranging from gene expression to motility and biofilm formation are constantly challenged by internal and external noise. While the importance of stochastic fluctuations has been appreciated for chemotaxis, it is currently believed that deterministic long-range fluid dynamical effects govern cell-cell and cell-surface scattering -- the elementary events that lead to swarming and collective swimming in active suspensions and to the formation of biofilms. Here, we report the first direct measurements of the bacterial flow field generated by individual swimming {\it Escherichia coli} both far from and near to a solid surface. These experiments allowed us to examine the relative importance of fluid dynamics and rotational diffusion 
for bacteria. For cell-cell interactions it is shown that thermal and intrinsic stochasticity drown the effects of long-range fluid dynamics, implying that physical interactions between bacteria are determined by steric collisions and near-field lubrication forces. This dominance of  short-range forces closely links collective motion in bacterial suspensions to self-organization in driven granular systems, assemblages of biofilaments, and animal flocks. For the scattering of bacteria with surfaces, long-range fluid dynamical interactions are also shown to be negligible before collisions; however, once the bacterium swims along the surface within a few microns after an aligning collision, hydrodynamic effects can contribute to the experimentally observed, long residence times. As these results are based on purely mechanical properties, they apply to a wide range of microorganisms.
\\
\\
Citation: Drescher K, Dunkel J, Cisneros LH, Ganguly S, Goldstein RE, \\
{\it Proc. Natl. Acad. Sci. (USA)} {\bf 108} 10940-10945.  \\
{\tt http://www.pnas.org/content/108/27/10940}
\end{abstract}

\maketitle

\begin{figure*}[t]
\centering
\includegraphics[clip=,width= 1.8 \columnwidth]{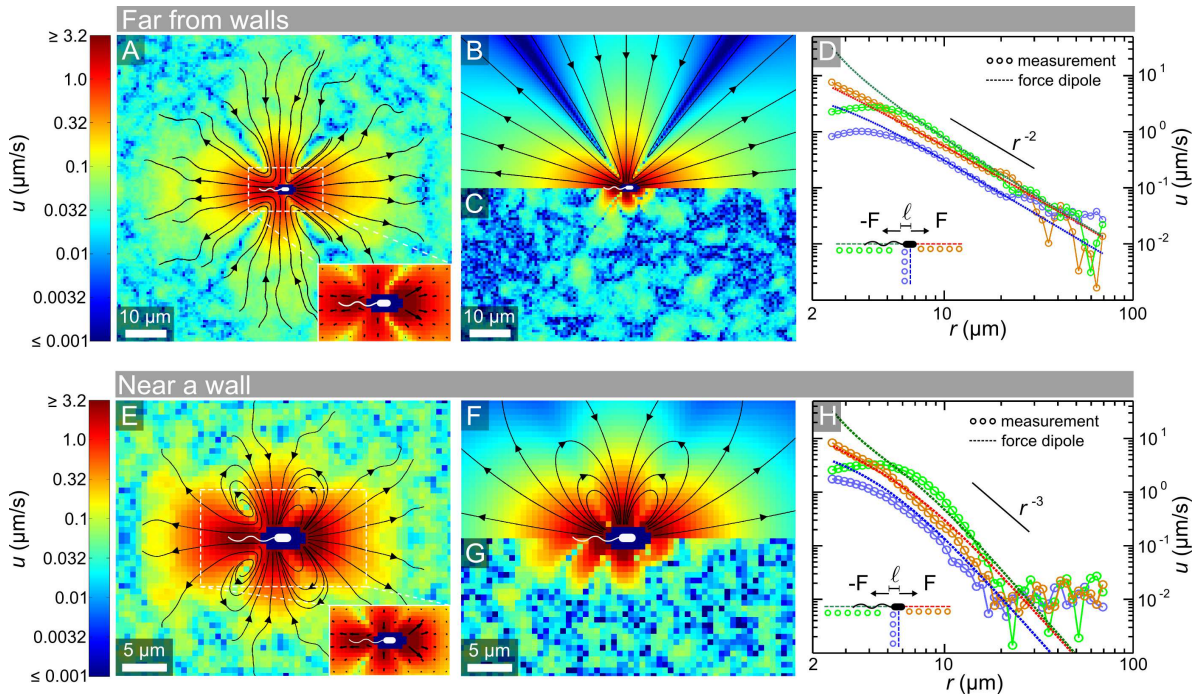}
\caption{\label{fig1} Average flow field created by a single freely swimming bacterium far from surfaces 
(A-D), and close to a wall (E-H). Streamlines indicate the local direction of flow, and the logarithmic 
color scheme indicates flow speed magnitudes.  (A) Experimentally measured flow field in the bacterial 
swimming plane, with the inset showing the anterior-posterior asymmetry close to the cell body. 
(B) Best-fit force dipole flow. (C) Residual flow field, obtained by subtracting the best-fit dipole 
model from the measured field. (D) Radial decay of the flow speed $u$ in different directions, 
with $r=0$ corresponding to the center of the cell body. For distances $r \lesssim 6$ $\mu$m the dipole model 
overestimates the flow field behind and to the side of the cell body. (E) Experimentally measured flow 
field in the bacterial swimming plane, 
for bacteria swimming parallel to a wall at a distance of $2$ $\mu$m. (F) Best-fit force dipole flow, 
where the presence of the wall causes inward and outward streamlines to join. (G) Residual flow field. 
(H) The flow speed decays much faster for bacteria swimming close to a wall, as the fluid velocity 
must vanish on the surface.}
\end{figure*}

Collective behavior of bacteria, such as biofilm formation \cite{Pratt}, 
swarming \cite{swarmingRev} and turbulence-like motion in concentrated suspensions 
\cite{Dombrowski,Wu}, has  profound effects on foraging, signaling, and transport 
of metabolites~\cite{Ecoli_BergInMotion,stockerPNAS}, and can be of great biomedical 
importance \cite{Donlan,biofilmsMedical}.  Large-scale coherence in bacterial systems 
typically arises from a combination of biochemical signaling~\cite{BLB} and physical 
interactions. Recent theoretical models that focus on physical aspects of  bacterial 
dynamics identify pairwise long-range hydrodynamic interactions 
\cite{Alexaner08,Guell88,Nasseri97,Ishikawa07,Najafi10,MatasNavarro10,Gyrya10} as a 
key ingredient for collective swimming. Such \lq\lq microscopic\rq\rq\space approaches 
underpin continuum theories that aim to describe the phenomenology of microbial suspensions~\cite{SimhaRama,SaintShelley,MarchettiPNAS,Graham,Aranson,IshikawaPedley,Putz10}. 
An assumption underlying many of these theories is that a self-propelled bacterium can 
be modeled  as a force dipole; its body exerts a drag force $F$ on the fluid that is 
balanced by the rearward flagellar thrust $-F$.  The leading order fluid velocity field 
at distance $r$ is therefore a dipolar \lq\lq pusher\rq\rq\space flow of magnitude 
$u\propto F \ell / \eta r^2$ (see streamlines in Fig.~\ref{fig1}B), where $\eta$ is the 
viscosity, and $\ell$ the distance between the forces \cite{Batchelor,LaugaPowers}. While 
higher order corrections may be due to force-quadrupole contributions  \cite{Liao}, the 
hypothesis that the leading-order flow field around a bacterium is dipolar has not yet 
been verified experimentally.

A closely related, controversially discussed issue \cite{Berke,Tang,elgeti,hernan} is the relevance of 
long-range hydrodynamic interactions in the scattering of bacteria with surfaces, a phenomenon 
intimately linked with surface accumulation and biofilm formation. Cell-surface scattering 
is very similar to cell-cell scattering, since, by analogy with image charges in electrostatics, a 
bacterium that swims near a surface induces an \lq\lq image bacterium\rq\rq\space on the opposite 
side of the wall to yield the no-slip boundary condition on the surface \cite{Berke,Blake}; 
bacterium-surface scattering can therefore be analyzed as the interaction of a bacterium with 
its hydrodynamic image. Several recent calculations for microswimmers near surfaces have found 
that pusher-type organisms (those with thrust generated behind the cell body) should display a 
passive stable alignment of the swimming direction with the wall \cite{Berke,Shum10,Goto05,Smith,Crowdy}. 
However, direct measurements of the three-dimensional swimming tracks of bacteria near surfaces suggest 
that they simply collide with the surface \cite{Tang,Frymier95}. This raises the question: Is this 
discrepancy between experiment and theory due to incomplete knowledge of the bacterial flow field, 
or is the magnitude of the flow so small that fluid-mediated interactions are irrelevant?

The need for experimental tests of the force dipole assumption and, more generally, of the relevance of 
fluid-mediated interactions for bacteria, is further illustrated by recent measurements on 
{\it Chlamydomonas reinhardtii}~\cite{PRL10,JandJ}, the archetypal 
\lq\lq puller\rq\rq\space microorganism (thrust generation in front of the cell body) that was 
thought to generate a simple force dipole flow with opposite sign to the bacterial one~\cite{LaugaPowers}. 
Surprisingly, these experiments showed that while such a dipolar flow exists at large distances from
the organism, in regions where the flow magnitudes are significant 
(more than 1\% of the swimming speed), the flow topology is qualitatively different, and more accurately
described in terms of a triplet of force singularities (one for the cell body and one for each
flagellum) \cite{PRL10}. Here we report the first direct measurement of the flow field around individual 
freely swimming bacteria, using {\it Escherichia coli} as a model. We find that the pusher force 
dipole provides a good approximation to the flow field both when the organism is far from surfaces and 
close to a no-slip boundary, yet the magnitude of the flow is very low. Using the experimentally determined 
flow field parameters, the hydrodynamic interaction of two {\it E. coli} can be calculated, and it 
is found to be washed out by rotational diffusion of the swimming direction for closest encounter 
distances~$\gtrsim 3$ $\mu$m -- a result that should hold for many other bacterial species due to the 
similarity of motility parameters. Similarly, analysis of cell-surface encounters suggests that  
hydrodynamics plays a negligible role, except when a bacterium swims along a surface at a small 
distance (less than a few microns) after an inelastic aligning collision. In this case, hydrodynamic 
effects can contribute to the observed long residence times near surfaces.

\section{Results}

\subsection{Flow Field Far from Surfaces}
To resolve the miniscule flow field created by bacteria, individual {\it gfp}-labeled, 
non-tumbling {\it E. coli} were tracked as they swam through a suspension of fluorescent 
tracer particles (see Materials and Methods). Measurements far from walls were obtained 
by focusing on a plane $50$ $\mu$m from the top and bottom surfaces of the sample chamber, 
and recording $\sim\! 2$ terabytes of movie data. Within this data we identified $\sim\! 10^4$ 
rare events when cells swam within the depth of field ($2$ $\mu$m thick) for $>1.5$ s. By 
tracking the fluid tracers during each of the rare events, relating their position and velocity 
to the position and orientation of the bacterium, and performing an ensemble average over all 
bacteria, the time-averaged flow field in the swimming plane was determined down to $0.1\%$ of 
the mean swimming speed $V_0 = 22 \pm 5$ $\mu$m/s. As {\it E. coli} rotate about their swimming 
direction, their time-averaged flow field in three dimensions is cylindrically symmetric. The 
present measurements capture all components of this cylindrically symmetric flow except the 
azimuthal flow due to the rotation of the cell about its body axis. In contrast with the flow 
around higher organisms such as {\it Chlamydomonas} \cite{PRL10,JandJ} and {\it Volvox} 
\cite{PRL10}, the topology of the measured bacterial flow field (Fig. 1A) is that of a force 
dipole (shown in Fig. 1B). Yet, there are some differences between the force dipole flow and 
the measurements close to the cell body, as shown by the residual of the fit (Fig. 1C). 

The decay of the flow speed with distance $r$ from the center of the cell body (Fig. 1D) illustrates 
that the measured flow field displays the characteristic $1/r^2$ form of a force dipole. However, the 
force dipole model significantly overestimates the flow to the side and behind the cell body, where the 
measured flow magnitude is nearly constant over the length of the flagellar bundle. The force 
dipole fit  to the far field ($r>8$ $\mu$m) was achieved with two opposite force monopoles (Stokeslets) 
at variable locations along the swimming direction. As $r=0$ corresponds to the center of the cell body in 
Fig. 1D, and not the half-way point between the two opposite Stokeslets, the fit captures some 
of the anterior-posterior asymmetry in the flow magnitude $u$. From the best fit, which is insensitive to the 
specific algorithms used, we obtained the dipole length $\ell = 1.9$ $\mu$m and dipole force $F = 0.42$ pN. 
This value of $F$ is consistent with optical trap measurements \cite{Chatto} and resistive force theory 
calculations \cite{Darnton}. It is interesting to note that in the best fit, the cell drag 
Stokeslet is located $0.1$ $\mu$m behind the center of the cell body, possibly reflecting 
the fluid drag on the flagellar bundle.

\subsection{Flow Field Near a Surface}
Having found that a force-dipole flow describes the measured flow around {\it E. coli} 
with good accuracy in the bulk (far from boundaries), we investigated whether this 
approximation is also valid when {\it E. coli} swim close to a wall. Focusing $2$ $\mu$m 
below the top of the sample chamber, and applying the same measurement technique as before, 
we obtained the flow field shown in Fig. 1E.  This flow decays much faster than that in 
the bulk due to the proximity of a no-slip surface (Fig. 1H), and the inward and outward 
streamlines are now joined to produce loops (Fig. 1E). However, both of these differences 
are consistent with a simple force dipole model, and are therefore not due to a change in 
bacterial behavior.  In particular, closed streamlines are known to be a rather general 
feature of point singularities near no-slip surfaces \cite{Pepper}. Using the solution of a 
Stokeslet near a wall \cite{Blake} to obtain that of a force dipole near a wall yields 
streamlines (Fig. 1F), and a decay (Fig. 1H) of the flow field that is consistent with the 
data. The best-fit force dipole in this geometry yields $F = 0.43$ pN and $\ell = 2.2$ $\mu$m, 
consistent with the values obtained far from walls, but less accurate than those values 
due to the much faster decay of $u(r)$ near a wall.

\subsection{Spectral Flow Analysis}
To analyze systematically the angular flow structure even at distances $r<6$ $\mu$m, where 
the force dipole model overestimates the flow magnitudes, it is useful to decompose the 
flow field into vector spherical harmonics. The resulting spectra are useful `fingerprints' 
of the flow field that can be compared among many different organisms, and against 
theoretical models. Such a spectral analysis is described in the SI Text.

\subsection{Rotational Diffusion}
Even bacteria that do not display tumbles, such as those studied here, do not swim in completely 
straight lines. Random changes in swimming direction due to thermal effects and intrinsic noise in the 
swimming apparatus lead to rotational diffusion which can be characterized by a coefficient $D_r$. 
From the swimming data recorded for {\it E. coli} far from surfaces, we measured $D_r = 0.057$ rad$^2$/s 
(see Materials and Methods). Even organisms that are too large to have significant thermal rotational 
diffusion, such as the 10-$\mu$m sized alga {\it Chlamydomonas reinhardtii}, can have a significant 
$D_r$ due to noise in the swimming mechanism \cite{Polinetal}. From swimming data previously recorded 
for {\it Chlamydomonas} \cite{PRL10}, we found $D_r = 0.4$ rad$^2$/s.

\section{Discussion}

Our measurements show that, independently of whether {\it E. coli} swim near or far from a 
surface, their flow field can be described to good accuracy by a simple force dipole model 
whose parameters we determined.  We now proceed to discuss the implications of this flow field for cell-cell 
and cell-surface interactions. Based on the measured parameters and the force dipole approximation, we 
calculate the effect of long-range hydrodynamics for these two scattering phenomena and 
evaluate the importance of fluctuations in the swimming direction.

\subsection{Hydrodynamics vs. Rotational Diffusion in Cell-Cell Scattering}

Fluid-mediated long-range interactions are thought to play an important role in 
collective motion in bacterial suspensions \cite{SimhaRama,SaintShelley,MarchettiPNAS,Graham,Aranson,IshikawaPedley,Putz10}. 
These deterministic interactions however compete 
with rotational diffusion of the swimming direction. To infer the importance of longe-range 
hydrodynamics in the bulk, we consider the change in swimming direction 
of a bacterium due to hydrodynamic scattering with another bacterium. This can be done by fixing one
bacterium at the origin and examining the trajectory of the other. The flow field around 
the bacterium at the origin is approximated by that of a point force dipole \cite{Blake}
\bea\label{e:dipole_model}
\bs u({\bs r})
=
\frac{A}{|\bs r|^2} \left[3(\bs{\hat r}\cdot{\bs{d}}')^2-1\right] \bs{\hat{r}}
,\quad
A=\frac{\ell F}{8\pi \eta}
,\quad 
\hat{{\bs r}}=\frac{\bs r}{|\bs r|},
\eea
where ${\bs{d}}'$ is the unit vector in the swimming direction, and $\bs r$ is now the 
distance vector relative to the center of the dipole. The evolution of the position ${\bs{x}}$ and 
swimming direction ${\bs{d}}$ of the second swimmer in this field obeys~\cite{1992PedleyKessler}
\bea
\bs{\dot{x}}
&=&
V_0 \bs{{d}}  +  \bs{u},
\label{e:wall_scattering_a}\\
\bs{\dot{{d}} }
&=&
\frac{1}{2} \; \bs{\go} \times \bs{{d}} +
\Gc\;
\bs{{d}}\cdot \bs{E}\cdot (\bs{I}-\bs{{d}}\bs{{d}})
\label{e:wall_scattering_b},
\eea
where $\bs{I}$ is identity matrix, and the central swimmer leads to an advective flow ${\bs u}$, 
a vorticity ${\bs \go}$, and a strain-rate tensor ${\bs E}$ at the position $\bs x$ 
(see SI Text for exact expressions of these quantities). By examining the evolution of 
${\bs{d}}(t)$ in a scattering process that begins at $-t/2$, reaches a minimal encounter 
distance $r$ at $t=0$, and ends at $t/2$, the mean squared angular change of orientation 
during a time interval $t$ can be computed as
\bea
\lan \Gd\phi(t,r)^2\ran_H =\left\lan\arccos[{\bs{d}}(-t/2)
\cdot{\bs{d}}(t/2)]^2\right\ran_H, 
\eea
where $\lan\;\cdot\; \ran_H$ indicates an average over all possible orientations and 
positions of encounters.  Assuming that the interaction time scale of the two bacteria 
$\tau$ is small, and using the force dipole model from Eq.~\eqref{e:dipole_model} we 
obtain (see derivation in SI)
\bea\label{e:hd_angle_change}
\lan \Gd \phi(\tau,r)^2 \ran_H
=
\frac{3}{5}(\Gamma + 1)^2  \frac{ A^2\tau ^2}{ r^6},
\eea
where $\Gamma \sim 1$ is a geometric factor for {\it E. coli}. Intuitively, this form
arises from the fact that $\Gd\phi\sim \omega \tau$, where the vorticity magnitude 
$\omega$ falls off as $A/r^3$. To evaluate the importance of hydrodynamic interactions 
relative to random fluctuations, we compare $\lan \Gd\phi(\tau,r)^2\ran_H$ with the 
angular diffusion due to Brownian motion and intrinsic swimming variability in three 
dimensions, given by $\lan \Gd\phi(t)^2\ran_D=4D_r t$. Balancing the effects of 
hydrodynamics and noise, 
\bea\label{e:criterion}
\lan \Gd\phi(\tau,r_H)^2\ran_H 
=
\lan \Gd\phi(\tau)^2\ran_D,
\eea
defines an effective hydrodynamic horizon $r_H$, beyond which noise dominates over 
hydrodynamics. For scattering events with closest encounter distances $r>r_H$ hydrodynamics 
is therefore practically irrelevant. From this definition of $r_H$ and Eq. 
\eqref{e:hd_angle_change}, we find
\bea
r_H \simeq \left[\frac{3}{20} (\Gamma +1)^2  \frac{
A^2\tau}{
D_r}\right]^{1/6}.
\label{eq:rhorizon}
\eea
Due to the $\tau^{1/6}$-dependence, the hydrodynamic horizon $r_H$ is rather insensitive 
to the particular value used for the interaction time scale $\tau$ and, similarly, to changes 
in the other parameters. Using the measured values $D_r = 0.057$ rad$^2$/s, $V_0=22$ $\mu$m/s, 
$A=31.8$ $\mu$m$^3$/s, and adopting  $\tau = a/V_0\simeq 0.1$ s, where $a = 3$ $\mu$m is the 
length of the cell body, we obtain $r_H\simeq 3.3$ $\mu$m for {\it E. coli}. This value of 
$r_H$ is an upper bound, as the dipolar flow model overestimates ${\bs u}$ for $r\lesssim 6$ $\mu$m (Fig. 1D). 
At such small separations, however, steric repulsion, flagellar intertwining, and lubrication forces 
dominate the physical cell-cell interactions \cite{GoldsteinPRE,Peruani}. For the mean distance 
between {\it E. coli} to reach $r_H$, the volume fraction needs to be as high as 
$5-10$\%.  Using our measured parameters in a recent theoretical calculation  \cite{Koch} of the critical 
volume fraction for the onset of collective swimming due to hydrodynamic interactions leads to an 
even higher value, implying that a complete analysis of collective behavior in bacterial 
suspensions must account for steric and near-field interactions.

More generally, we expect similar results to hold for various types of swimming bacteria 
(e.g., {\it Pseudomonas aeruginosa}, {\it Vibrio cholerae}, {\it Salmonella typhimurium}), as 
the parameters $(a,\tau, D_r, F \ell)$ are similar across many genera.  Larger organisms may display 
even stronger rotational diffusion due to enhanced intrinsic swimming stochasticity \cite{Polinetal}.  
For example, the alga {\it Chlamydomonas} ($a \simeq 10$ $\mu$m, $V_0 \simeq 100$ $\mu$m/s) has 
$D_r \simeq 0.4$ rad$^2$/s. Although the flow topology around {\it Chlamydomonas} is more 
complicated than that around bacteria, {\it Chlamydomonas} still produces a $1/r^2$ field \cite{PRL10}, 
so that our previously calculated result for the bacterial hydrodynamic horizon may be used to give 
an estimate of $r_H \sim 7.5$ $\mu$m, again on the scale of the organism. Thus, collisions, rather 
than long-range hydrodynamics, can also govern scattering events of higher microorganisms. However, 
for organisms that produce fast flows, have a long interaction time $\tau$, and a negligible $D_r$, 
like the alga {\it Volvox}, long-range hydrodynamic interactions are significant \cite{PRL09}.

\subsection{Cell-Surface Scattering}

The accumulation of bacteria near surfaces is a key step during the initial stages of biofilm formation, 
and it has been suggested that long-range hydrodynamics plays an important role in 
bacteria-surface interactions \cite{Berke}.  As our measurements show that the field has a 
small amplitude, it is plausible that bacterial cell-surface scattering events could be described instead by 
nearly straight-line swimming interrupted by collisions with the wall which lead to alignment with the surface 
due to near-field lubrication and/or steric forces during the collision \cite{Tang}. 
Our experimental results establish the key microscopic parameters required for a systematic
investigation of whether long-range hydrodynamic interactions are relevant to
bacteria-surface scattering. 

\begin{figure}[t!]
\centering
\includegraphics[width=1.00\columnwidth]{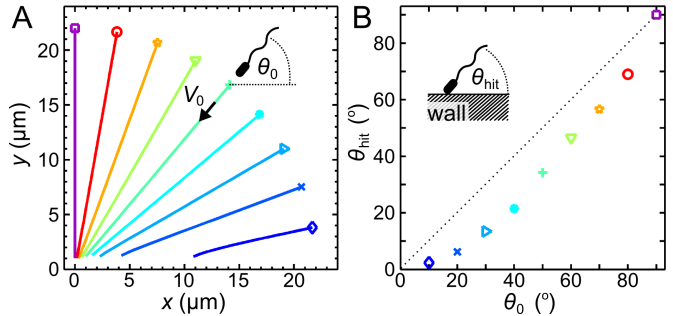}
\caption{
\label{fig:wall_collision} Simulated dynamics of an {\it E. coli}-like force-dipole swimmer near a wall. 
(A) Deterministic swimming trajectories towards a wall at $y=0$, numerically simulated 
from Eqs.~\eqref{e:wall_scattering_a} and \eqref{e:wall_scattering_b}, where $\bs u$, 
$\bs{\omega}$, and $\bs E$ are due to the hydrodynamic image system. Simulations used a time step 
$\Gd t=10^{-5}\sec$ and the experimentally 
determined parameters $A=31.8$ $\mu\met^3/\sec$, $V_0=22$ $\mu\met/\sec $ and $\Gc= 0.88$ for the 
force-dipole swimmer. The initial distance is chosen such that the swimmer would reach the wall 
after $1$ s if hydrodynamic interactions were not present. (B) Incidence angle $\gt_0$ \textit{vs.} 
collision angle $\gt_\mrm{hit}$ with the wall for the trajectories in panel A, using the same 
symbols and colors. The dotted line indicates $\gt_\mrm{hit} = \gt_0$. Both panels illustrate 
that hydrodynamic long-range interactions can be regarded as small perturbations 
for typical wall scattering events of {\it E. coli}.}
\end{figure}

As a first step in our analysis, we performed numerical scattering studies by 
simulating the deterministic equations of motion for an {\it E. coli}-like pusher 
force-dipole swimmer in the presence of an infinite no-slip surface. The equations 
of motion of the swimmer position~$\bs x$ and unit orientation vector $\bs{{d}}$ 
are simply those used above (Eqs. \eqref{e:wall_scattering_a} and 
\eqref{e:wall_scattering_b}) except that now $\bs{u}$, ${\bs{\go}}$, and 
${\bs{E}}$ are quantities arising from the hydrodynamic image system in the wall 
(exact expressions for these quantities are given in the SI Text). By restricting 
ourselves to simulations of the deterministic dynamics at this stage, we overestimate 
the relevance of hydrodynamic long-range interactions between the swimmer and the 
wall, as rotational diffusion of the swimming direction further diminishes 
hydrodynamic effects. However, even without rotational diffusion our simulations 
show that long-range interactions of swimmers with the wall have little effect on 
the swimming dynamics (Fig.~\ref{fig:wall_collision}). The trajectories of 
force-dipole swimmers that swim towards the wall from different initial angles 
$\theta_0$ are depicted in Fig.~\ref{fig:wall_collision}A. The initial distance was 
chosen such that the swimmer would reach the wall plane ($y=0$) after $1$~s if 
hydrodynamic interactions were absent. Each simulation is stopped when a volume 
around the swimmer (a bacterial shape of length 3 $\mu$m and diameter 0.8 $\mu$m) 
crosses the wall. Figure~\ref{fig:wall_collision}B displays the impact angle 
$\gt_\mrm{hit}$ as a function of the initial angle $\gt_0$, illustrating that the 
difference between incidence and collision angles is small unless the swimmer 
already has a small angle of incidence. These simulations indicate that hydrodynamic 
long-range interactions are not likely to play an important role in cell-surface 
scattering for {\it E. coli}. As the swimming parameters are similar for many 
bacterial genera, we again expect this result to apply more generally.

\subsection{Trapping by Surfaces}
When {\it E. coli} swim very close to a surface ($\sim 1-3$ $\mu$m), we observed that 
individual bacteria spend an average of $64$ s (standard error $4$ s) next to the wall. 
Effective trapping by electrostatic attraction is unlikely, since both the {\it E. coli} 
outer cell wall and the chamber walls (bovine serum albumin coated onto PDMS \cite{BSApaper}) 
are negatively charged  in our liquid medium (we observed similarly long residence times 
on simple glass surfaces). However, the surprisingly long residence times could be caused 
by the suppression of rotational diffusion due to geometric constraints on the orientation 
of the cell body and flagella near a surface. Although we showed in the previous section 
that hydrodynamics has a very small effect on the swimming direction before collisions with 
the surface, hydrodynamic attraction by the surface \cite{Berke} could contribute to the 
observed trapping periods when a bacterium is already very close to the surface. Considering 
only hydrodynamic attraction counteracted by rotational diffusion, we now derive approximate 
expressions for the mean escape time $t_e$ and escape height above the surface $h_e$, by 
mapping the underlying escape process onto a Kramers' problem~\cite{HanggiReview,DunkelKramers}  
for the noise-induced escape over a potential barrier. The main arguments and implications 
are summarized below, while a detailed derivation is given in the SI Text.

We again approximate the {\it E. coli} flow field 
by the dipole model, because a force-dipole placed close to a wall accurately 
captures the measured flow field parallel to the surface (see Fig. 1E-H). Thus, 
Eq. \eqref{e:dipole_model} is modified to account for the presence of the wall 
\cite{Blake}, as discussed in the SI Text, and near-field hydrodynamic lubrication 
effects are neglected. A bacterium is able to escape from the surface, if its 
swimming velocity component perpendicular to the surface, $V_0\sin\gt$, exceeds the 
attraction from its hydrodynamic image (see SI Text), which yields the
defining relation for the escape angle $\gt_e$,
\bea\label{e:escape_angle}
\sin\gt_e=\Gl \left[ 1-3 (\sin\gt_e)^2\right],
\qquad
\Lambda=\frac{3A}{8h^2V_0}.
\eea
For \textit{E. coli} swimming at distances $h > 1.5$ $\mu$m from the wall, the escape 
angles are small, $\gt_e(h) < 11^{\circ} \ll 1$ rad so that linearization of 
Eq.~\eqref{e:escape_angle} is a good approximation, giving $\gt_e\simeq \Gl$.

After colliding with the wall, a bacterium may have a small positive angle  $\gt<\gt_e$ 
with the surface. The equation of motion for $\gt$ can then be rewritten as a Langevin 
equation \cite{HanggiReview} in terms of the derivative of an effective `potential' 
$U(\gt)$, and a diffusion term with Gaussian white noise $\xi(t)$,
\be
\dot\gt
=-\frac{\diff U}{\diff \theta}
+ (2D^*_r)^{1/2} \xi(t) \:,
\qquad
U(\gt) \simeq \frac{\gt^2}{2\kappa} \: ,
\label{e:Langevin}
\ee
where the approximation $\theta \ll 1$ reduces $U(\theta)$ to a 
harmonic potential, yielding a time scale $\kappa = 16 h^3 / (9 A)$ that characterizes
hydrodynamic realignment. $D^*_r$ is the rotational diffusion constant close to 
the surface in the direction perpendicular to the surface, which is expected to be 
smaller than our measured value $D_r=0.057$~rad$^2$/s far from boundaries, 
due to geometric constraints on the bacterial orientation near a surface. 
The generic form of this Langevin equation means that finding the residence time for a 
bacterium near a wall is a Kramers problem~\cite{HanggiReview,DunkelKramers} 
for the escape over a barrier $\Gd U$. As the organism can escape if $\gt>\gt_e$, 
we have  $\Gd U=U(\gt_e)$.

By considering the height at which $\Gd U = D^*_r$, i.e., the distance at which 
hydrodynamic torque is comparable to diffusion, we can obtain an expression for 
the escape height 
\be
h_e = \frac{1}{2} \left( \frac{81}{16} \frac{A^3}{D^*_r V^2_0} \right)^{1/7}.
\label{equ:escapeheight}
\ee
Using our measured values for {\it E. coli}, we find
$h_e = 1.7 \: \mu\mbox{m} \times (D_r / D^*_r)^{1/7}$, illustrating that hydrodynamics is 
practially negligible if {\it E. coli} are more than a cell length away from the wall.

As long as the torque exerted by the hydrodynamic image is small, $\Delta U \ll D^*_r$, 
the typical escape time is set by the rotational 
diffusion time scale $\theta^2_e / D^*_r$. For high barriers 
$\Delta U \gg D^*_r$ (in practice, $\Delta U > 3 D^*_r$ often suffices, 
yielding $h \lesssim 1.5\times (D_r / D^*_r)^{1/7}\; \mu$m), transition state theory 
\cite{HanggiReview} implies that the mean escape time is modified by an 
Arrhenius-Kramers factor, so that approximately
\be
t_e(h) \approx \left( \frac{\theta_e^2}{D^*_r} \right) \: 
\exp \left( \frac{\Delta U}{D^*_r} \right).
\label{equ:escapet1}
\ee
Using the quadratic approximation for $\Delta U$, and Eq. \eqref{equ:escapeheight} to 
express $D^*_r$ in terms of $h_e$, we find that $t_e \propto \exp (h_e / h)^7$. 
This dramatic scaling arises from the fact that the dipole model overestimates 
the flow field close to the bacterium, but generally hints at the possibility of a 
strong, hydrodynamically induced increase of $t_e$ when the cells get closer to 
the surface. We may also evaluate Eq. \eqref{equ:escapet1} at a 
height $h= 1.5$~$\mu$m, where both the Arrhenius-Kramers factor and the approximation 
$\theta_e(h) \ll 1$ are valid, to give
\be
t_e \approx 0.78 \: \mbox{s} \times \left( \frac{D_r}{D^*_r} \right) 
\exp \left[ 1.99 \times \left( \frac{D_r}{D^*_r} \right) \right].
\ee  
The latter estimate suggests that hydrodynamic effects can possibly explain 
the experimentally observed long residence times near a wall, even for values of 
$D_r / D^*_r$ that are only moderately larger than $1$. 
It is, however, important to note that this expression for $t_e$ presents an upper bound, 
because the dipole model overestimates the actual flow field 
at distances $<6$~$\mu$m from the bacterium (Fig. 1H), even though the model 
still correctly captures the flow topology.

The above considerations show that hydrodynamics is negligible if a bacterium is more 
than a body length away from the wall, but that hydrodynamic effects may contribute to 
the experimentally observed long residence times of bacteria close to no-slip surfaces. 
A more detailed understanding of the escape problem remains an important future challenge, 
requiring new methods for measuring $D^*_r$ and further theoretical studies of the near-field 
interactions between bacteria, their flagella, and surfaces. However, even if a more 
accurate description of the hydrodynamics should become available in the future, one can 
still expect the mean escape time to follow an Arrhenius-Kramers law of the form~\eqref{equ:escapet1} 
with a suitably adapted effective potential $U$ and additional prefactors that account 
for the curvature at the barrier~\cite{HanggiReview}.

\section{Conclusions}

We have presented the first direct measurement of the flow field generated by individual freely-swimming 
bacteria, both in bulk fluid and close to a solid surface. For distances $\gtrsim 6\,\mu$m, the experimentally 
measured flow field is well-approximated by a force-dipole model; at smaller distances the 
dipole model overestimates the flow. Generally, the flow field of {\it E.~coli} differs markedly from 
those created by higher microorganisms, such as {\it Chlamydomonas} \cite{PRL10,JandJ} and 
{\it Volvox} \cite{PRL10}. With regard to the future classification of flow fields of microorganisms, 
a decomposition in terms of vector spherical harmonics can provide a useful systematic 
framework, similar to the classification of the electronic orbital structures in atoms or molecules. 

Theories of collective behavior in bacterial suspensions identify as a fundamental process the 
pairwise interaction of bacteria, often assumed to be dominated by long-range fluid flows established 
by the action of swimming \cite{LaugaPowers}.  Our analysis suggests that noise, due to orientational 
Brownian motion and intrinsic swimming stochasticity, drowns out hydrodynamic effects between 
two bacteria beyond a surprisingly small length scale of a few microns. This implies that hydrodynamic 
effects will be relevant only in sufficiently dense bacterial suspensions. However, under such conditions, 
the flow structure close to the bacterial body and contact interactions (e.g., flagellar bundling, 
steric repulsion) will be more important than the asymptotic long-range details of individual 
microswimmer flow fields. 

Insights into the biochemical and physical interactions between bacteria and surfaces are crucial for 
understanding the dynamics of biofilm formation, the emergence of collective bacterial behavior in 
boundary layers and, thus, more generally the evolution from unicellular to multicellular, 
cooperative forms of life. Our results suggest that long-range hydrodynamic effects play a 
negligible role in the scattering of {\it E. coli} with surfaces before collisions. 
However, hydrodynamic effects can, at least partially, account for the observed trapping of bacteria within 
a few microns of the surface. The analysis presented herein lends support to the hypothesis 
\cite{ZBN} that turbulent swarming patterns in bacterial films arise primarily due to 
steric repulsion and other near field interactions, since long-range hydrodynamic interactions 
become suppressed near surfaces due to cancellation effects from the image swimmer. 

Our experimental and theoretical results favor collision-dominated models \cite{Tang,elgeti,hernan} 
for the accumulation of bacteria at surfaces over models based on long-range hydrodynamics~\cite{Berke}. 
To obtain a more complete dynamical picture of biofilm formation, future efforts should focus on 
developing more precise measurement methods and advanced models that include lubrication effects 
and biochemical bacteria-surface interactions. While our combination of measurements, simulations, 
and theory shows that long-range physical interactions are negligible for bacterial cell-surface 
scattering, fluid-mediated coupling could become important for organisms swimming against or in 
contact with a surface, as the organism is then no longer force-free, resulting in a substantially 
longer range of hydrodynamic interactions \cite{PRL09,Cortez}. 

However, the main implication of the present study is that short-range forces are likely to 
dominate the interactions between swimming bacteria, so that collective motion in bacterial 
suspensions, thin films \cite{Wu,SokolovPRL} and thin wetting layers \cite{Swinney} relates 
closely to that seen in driven granular systems \cite{Granular}, 
assemblages of biofilaments \cite{biofilaments}, and animal flocks 
\cite{RamaswamyColMot,CouzinColMot}. This suggests that many of the principles 
that determine flocking and self-organization in higher animals should also govern the collective 
motion of the smallest organisms.

\section{Materials and Methods}
A detailed description of the mathematical models is provided in the SI Text. The 
experiments are summarized below.

\subsection{Culture Conditions}
We used {\textit{Escherichia coli}} strain HCB437 carrying the plasmid pEGFP (Clontech, BD Biosciences), 
kindly supplied by Douglas B. Weibel (University of Wisconsin-Madison) and Howard C. Berg (Harvard 
University). Cells were streaked on 1.5$\%$ agar plates containing T-broth (1$\%$ tryptone, 0.5$\%$ NaCl) 
and $100$ $\mu$g/mL ampicillin. A single-colony isolate from an overnight plate was used 
to innoculate $10$ mL of T-broth containing ampicillin and $0.1$ mM isopropyl 
$\beta$-D-1-thiogalactopyranoside (IPTG, Sigma), which was then grown for $7$ h on a rotary 
shaker ($200$ rpm) at $33~^{\circ}$C. This culture was diluted 1:1 with fresh T-broth 
containing ampicillin and IPTG as above, $0.2\%$ bovine serum albumin, and $0.2$ 
$\mu$m fluorescent microspheres (505/515, F8811, Invitrogen) at concentration 
$9\times 10^9$ beads/mL. This bacterial suspension ($\sim 1.6 \times 10^7$ cells/mL) was loaded 
into a polydimethylsiloxane (PDMS) microfluidic device consisting of cylindrical measurement chambers 
(height 100 $\mu$m, radius $750$ $\mu$m) connected by thin channels. After filling the device, it 
was sealed to reduce background fluid motion.

\subsection{Measurement of the Flow Field}
Using a Zeiss Axiovert inverted microscope with a 40$\times$ oil objective (NA 1.3), we 
simultaneously imaged bacteria and microspheres under fluorescence conditions at $40$ fps 
(Pike, Allied Vision Technologies) and at a temperature of $24 \pm 1$~$^{\circ}$C. To measure 
the flow field far from walls, we focused on a plane $50$ $\mu$m  inside the chamber to minimize 
surface effects. To measure the flow field close to a no-slip surface, we focused on a plane 
$2$ $\mu$m below the top surface of the sample chamber. Each movie was analyzed with 
custom Matlab software that precisely tracked bacteria by fitting an ellipsoidal two-dimensional 
Gaussian shape. For each cell swimming along the focal plane 
for $>1.5$ s, we collected the instantaneous velocity of all fluorescent tracers up to a 
distance of $75$ $\mu$m, using standard particle tracking algorithms. The resulting 
$\sim 5 \times 10^9$ tracer velocity vectors were binned into a $0.63$ $\mu$m square grid 
(shown in Fig. 1A, E). The mean of the well-resolved Gaussian distribution in each bin was 
taken as a local measure of the flow field. To measure the mean residence time of bacteria near 
a surface, we used the movies that were recorded for measuring the flow field near the 
wall. 

\subsection{Measurement of the Rotational Diffusion} 
From the tracks of {\it E.~coli} that swam in the focal plane for $>1.5$ s, at a distance of 
$50$ $\mu$m from the top and bottom surfaces, we determined an average swimming direction at 
time $t$ by using the direction between the bacterial positions at $t-0.05$ s and $t+0.05$ s. 
Computing the change in average swimming direction $\Delta \phi$ revealed diffusive scaling, 
so that we obtained $D_r$ from the equation for two-dimensional orientational diffusion, 
$\left\langle | \Delta \phi | ^2 \right\rangle = 2 D_r \Delta t$, over a time interval $\Delta t$. 
We measured $D_r$ for {\it Chlamydomonas} with the same procedure, using cell-tracking data described 
earlier \cite{PRL10}.

\section{Acknowledgments}
We are indebted to H.C. Berg and D.B. Weibel for providing bacterial strains and advice, and
thank I.S. Aranson, P. H\"anggi, S. Hilbert, T.J. Pedley, P. Talkner, and I. Tuval 
for discussions, and the reviewers for their comments. This work was supported in part 
by the Engineering and Physical Sciences Research Council, the Biotechnology and Biological 
Sciences Research Council, and the European Research Council.

\end{document}